\documentclass[preprint,tightenlines,aps,floats,nofootinbib]{revtex4}

\usepackage{epsf}
\usepackage{epsfig}
\usepackage{graphicx}
\usepackage[dvips]{color}
\usepackage{amsmath,amssymb,bbm,mathrsfs}

\newcommand{\notE}{\ \hbox{{$E$}\kern-.60em\hbox{/}}}
\newcommand{\notp}{\ \hbox{{$p$}\kern-.43em\hbox{/}}}
\def\D0{\mbox{D\O}}
\def\eslt{E_T^{\rm miss}}
\newcommand\jhep[3]{{\it J. High Energy Phys.\ }{\bf #1} (#2) #3}
\newcommand\plb[3]{{Phys.\ Lett.\ }{\bf B #1} (#2) #3}
\newcommand\npb[3]{{Nucl.\ Phys.\ }{\bf B #1} (#2) #3}
\renewcommand\prd[3]{{Phys.\ Rev.\ }{\bf D #1} (#2) #3}

\newcommand{\eps}{\epsilon}

\includegraphics{ou-seal.eps}

\preprint{\font\fortssbx=cmssbx10 scaled \magstep2
\hbox to \hsize{
\hskip1.2in 
\hbox{\fortssbx The University of Oklahoma}
\hskip0.2in $\vcenter{
                      \hbox{\bf OUHEP-111212}
                      \hbox{\bf arXiv: [hep-ph]}
                      \hbox{December 2011}}$ }
}

\begin{document}
 
\title{\vspace*{0.7in}
Prospects for Higgs Searches with the Tri-bottom Channel in 
Unified SUSY Models}

\author{
Howard Baer\thanks{Email address:baer@physics.ou.edu},
Chung Kao\thanks{Email address: kao@physics.ou.edu}, and 
Joshua Sayre\thanks{Email address: sayre@physics.ou.edu}
}

\affiliation{
Homer L. Dodge Department of Physics and Astronomy \\
University of Oklahoma \\ 
Norman, Oklahoma 73019, USA 
\vspace*{.4in}}

\date{\today}

\begin{abstract}

We investigate the prospects for the discovery of a neutral Higgs
boson produced in association with a $b$ quark, 
followed by the Higgs decay into a pair of bottom quarks, 
$pp \to b\phi^0 \to b b\bar{b} +X$,  
at the CERN Large Hadron Collider (LHC) within the framework of 
unified supersymmetric models.
The Higgs boson $\phi^0$ can be a heavy scalar $H^0$ or a pseudoscalar $A^0$.
Furthermore, this direct discovery channel is compared with the indirect Higgs
searches in the rare decay $B_s \to \mu^+\mu^-$ at hadron colliders. 
Promising results are found for the minimal supergravity (mSUGRA) model, 
the anomaly mediated supersymmetry breaking (AMSB) model, and 
the gauge mediated supersymmetry breaking (GMSB) model.
We find that the indirect search for 
$B(B_s \to \mu^+\mu^-) \ge 5\times 10^{-9}$ is complementary 
to the direct search for $b\phi^0 \to bb\bar{b}$ with $\sqrt{s} = 14$ TeV 
and an integrated luminosity ($L$) of 300 fb$^{-1}$.
In the AMSB and GMSB models, $b\phi^0 \to bb\bar{b}$ with $L = 300$ 
fb$^{-1}$ covers a larger area in the parameter space than 
$B(B_s \to \mu^+\mu^-) \ge 5\times 10^{-9}$.
In addition, we present constraints from $b \to s\gamma$ and muon 
anomalous dipole moment ($\Delta a_\mu$) on the parameter space.

\end{abstract}

\pacs{PACS numbers: 14.80.Da, 12.10.Dm, 12.60.Jv, 13.85.Ni}
%
%

\maketitle


\color{black}

\section{ Introduction}

Although the Standard Model (SM) remains an incredibly successful description of most of the 
phenomena studied by high energy physics, it is expected to be replaced by a more complete
model in order to address a number of theoretical problems. There are also several experimental 
signatures for which the Standard Model does not seem to account. Perhaps the most pressing theoretical 
question is the stabilization of the hierarchy between the weak and Planck scales. Supersymmetry (SUSY) 
provides the most popular solution to this problem and, due to the large number of exotic particles introduced,
also suggests solutions to a variety of other Standard Model anomalies. These include the origin of dark matter
and deviations in flavor and CP physics. On the other hand, SUSY must be a broken symmetry at low scales and 
in the absence of a detailed theory of spontaneous breaking we are left with a large number of free parameters 
in the form of soft supersymmetry breaking terms. To ameliorate this profusion of parameters, it is useful to 
adopt one of several unified frameworks which reduce the number of new variables. 

For phenomenological
reasons, it is assumed that supersymmetry breaking is driven by the dynamics of a hidden sector separate from
the Standard Model, and this breaking is communicated to the Standard Model particles and superpartners by a 
messenger sector which couples to both.  Gravity will act as a natural candidate for this messenger, leading to supergravity 
theories (SUGRA). However, it is possible that the gravity-induced SUSY breaking terms are not dominant and 
that the leading effects are generated either by messenger fields which carry Standard Model gauge numbers 
(gauge-mediated supersymmetry breaking, or GMSB), or by terms arising from the superconformal anomaly 
(anomaly-mediated supersymmetry breaking, or AMSB). 

It is also a striking fact that adding the minimal particle content required to 
incorporate the SM fields in a supersymmetric theory changes the running of the gauge couplings such that
they unify at a high scale (approximately $10^{16}$ GeV). This is a much better unification than is seen for the
SM fields alone and is consistent with the Grand Unified Theory (GUT) hypothesis that the three Standard Model 
gauge groups are remnants of a larger symmetry which is restored at high scales.

With the assumptions of grand unification and the messenger sector for SUSY breaking, the theory can be 
characterized by measured Standard Model inputs and a small number of additional unification scale parameters.
This is enough to determine all the masses and couplings of all the superpartners, allowing us to predict the 
potential for direct discovery and to calculate indirect effects, which appear as corrections to precision measurements 
relative to their Standard Model values. 

The LHC is rapidly accumulating data, currently with $\sqrt{s} = 7$ TeV, and already setting new limits on 
observables sensitive to new physics. It is planned to run at this energy through 2012, followed by a long 
shutdown for refurbishing and then a new run in 2014 aiming at the design energy of $\sqrt{s} = 14$ TeV. With 
high energy and luminosity, the LHC should be able to discover both the Higgs boson and supersymmetry if they 
exist in the $100-1000$ GeV range as anticipated by theorists. 

One of the basic predictions of supersymmetry is the existence of (at least) two Higgs doublets, in contrast with
the Standard Model. This leads to five physical states at low energy: two neutral scalars $h^0$ (lighter) and $H^0$ (heavier), a 
neutral pseudoscalar $A^0$, and a pair of charged scalars $H^{\pm}$. Thus a key signature of supersymmetry would
be the detection of this extended Higgs sector. In this paper, we focus on the prospects for detecting the heavy 
neutral Higgs scalar and the Higgs pseudoscalar. We investigate the
decay channel $\phi^0 \to b\overline{b}, \phi^0 = H^0, A^0$, 
which becomes important for large values
of the ratio of vacuum expectation values (VEVs) ($\tan\beta \equiv v_2/v_1$). 

We have considered this channel in a previous paper in the context of
the minimal supersymmetric standard model (MSSM) \cite{hbbb}. In this
study, we choose instead to work in the three unifying frameworks mentioned above. We plot the 
estimated detection contour in the space of the unification scale parameters for mSUGRA, mGMSB, and mAMSB 
respectively. These models also make predictions for rare decay rates and precision parameters which are currently
measured or constrained. In particular, we compare the LHC discovery
potential of the $3b$ channel with $B_s \to \mu^+ \mu^-$ at hadron
colliders, which is especially enhanced by a large value of $\tan\beta$. 
Furthermore, we consider constraints from $b \to s\gamma$ and 
and the anomalous magnetic moment of the muon, $(g-2)_{\mu}$. 

In our mSUGRA and mAMSB plots, the lightest neutralino $\tilde{\chi}$ is always required to be
the lightest SUSY particle, or LSP.\footnote{In mGMSB, the gravitino $\tilde{G}$ is taken as LSP.} 
If $R$-parity is conserved, 
then the $\tilde{\chi}$ will be absolutely stable and may comprise all or some of the 
cold dark matter (CDM) in the universe. 
Within the confines of the MSSM, a neutralino as CDM requires rather high 
fine-tuning\cite{bbox} and also very low reheat temperature $T_R<10^5$ GeV
in order to avoid bounds on late-decaying gravitinos from BBN\cite{bbn}.
If we invoke late-decaying scalar fields, such as TeV-scale moduli, into our theory, 
then these can either increase or decrease the standard neutralino abundance\cite{gondolo}.
Alternatively, invoking the Peccei-Quinn solution to the strong CP problem
then requires either mixed axion/axino\cite{bbs} or mixed axion/neutralino
CDM\cite{az1}, for which again the relic density predictions can be quite different.
Here, we will simply refrain from implementing any such constraints.

\section{ Higgs Production in Association with b Quarks}

The Minimal Supersymmetric extension of the Standard Model (MSSM) requires two Higgs doublets with opposite 
hypercharge to account for fermion masses since the superpotential must be analytic: 
\begin{align}
  \phi_1 = \begin{pmatrix}\phi_1^+ \\ \phi_1^0 \end{pmatrix} \, ,
  \qquad   \phi_2 = \begin{pmatrix} \phi_2^0 \\ \phi_2^- \end{pmatrix} \, .
\end{align}

Each doublet acquires a VEV in its neutral component, $v_1$ and $v_2$. The model is characterized by the ratio of 
these VEVs $\tan \beta \equiv v_2/v_1$. After symmetry breaking three of the degrees of freedom between the 
doublets are eaten by the massive weak vector bosons and the remaining five become the massive physical 
eigenstates $h^0,H^0,A^0$ and $H^{\pm}$ which are a mixture of the original doublet fields. By convention, 
$h^0$ is lighter than $H^0$. We will use $\phi^0$ to refer generically to any of the neutral Higgs bosons.

At the LHC, $\phi^0$ is predominantly produced by gluon fusion ($gg \to \phi^0$) for $\tan \beta \lesssim 5$. 
However, in a 2HDM the coupling of down-type quarks to the Higgs is proportional to the quark mass divided by $\cos \beta$,
which can be quite large compared to the Standard Model value. Hence for $\tan \beta \gtrsim 7$ the leading 
production of $\phi^0$ comes from $b\overline{b}$ fusion~\cite{
Dicus1,Dicus2,Balazs,Maltoni,Harlander}. 
Similarly, the branching fraction for $\phi^0 \to b\overline{b}$ can
become nearly $90\%$. 

In the case of high $\tan \beta$, the simplest channel to look for neutral Higgs would seem to be $b\overline{b} 
\to \phi^0 \to b\overline{b}$. However, this signal is swamped by large QCD dijet backgrounds. The next step is to 
take advantage of b-tagging capabilities by looking for $\phi^0$ in association with one or more additional b-jets. 
Here we have the option of requiring four b-tagged jets with high
$p_T$, $pp \to \phi^0 b \overline{b} \to bb\overline{bb} +X$, 
which will greatly 
reduce the background compared to the dijet case but which suffers from a small signal in turn~\cite{Vega,Yuan,Carena:1998gk}.
 In this paper, 
we will pursue the intermediate case with 3 high-$p_T$ tagged b-jets in the final state. It has been argued that this channel
is more promising at the LHC \cite{Scott2002}. 

We treat this channel in a 5-flavor quark Parton Distribution Function (PDF) scheme. In a 4-flavor scheme, the leading order 
process of interest would be $gg \to \phi^0 b\overline{b}$, where the Higgs
 subsequently decays
back to b-jets. One would then need to sum over all configurations with 3 high-$p_T$ quarks in the final state. This requires
careful treatment of higher-order corrections since the 4th quark, at low $p_T$, gives rise to large leading-log corrections. In
the 5-flavor scheme the b-quark is treated as an initial parton and the large corrections are absorbed by the b-quark PDF. For
5 flavors the leading order process is $bg \to b\phi^0 \to bb\overline{b}$. (As well as the conjugate process with 
$\overline{b}\phi^0$. Henceforth we will use $b$ to refer to both $b$ and anti-$b$ quarks.) The process $bg \to b\phi^0$ has
two Feynman graphs and by choosing the factorization and renormalization scale $\mu_F = \mu_R = M_H/4$, NLO corrections can be made relatively small \cite{Maltoni,Scott2002}. 

The decay of the Higgs boson $\phi^0 \to b\bar{b}$ depends at first order on the size of the Yukawa coupling to bottom quarks, 
which scales with $\tan \beta$. Here, we take into account several higher order effects since we wish the effective size of
this vertex to correspond with the Higgs decay width to $b\overline{b}$. The decay width at NLO can be written \cite{Guasch:2003cv}
\begin{align}
\Gamma(\phi^0 \to b\overline{b}) = \frac{3 G_F M_{\phi}}{4\sqrt{2}\pi}\overline{m}_b^2(M_\phi)[1+\Delta_{QCD} + \Delta_{t}] (g_b^{\phi})^2,
\end{align}
where
\begin{align}
\Delta_{QCD} = 5.67\frac{\alpha_s(M_\phi)}{\pi} +(35.94 - 1.36N_F)(\frac{\alpha_s(M_\phi)}{\pi})^2 \\
\Delta_t^A = \cot^2 \beta[3.83-\ln(\frac{M_A^2}{M_t^2})+\frac{1}{6}\ln(\frac{M_A^2}{M_t^2}))^2](\frac{\alpha_s(M_\phi)}{\pi})^2 \\
\Delta_t^{h}=-\cot \alpha \cot \beta[1.57-\frac{2}{3}\ln(\frac{M_h^2}{M_t^2})+\frac{1}{9}\ln(\frac{M_A^2}{M_t^2}))^2](\frac{\alpha_s(M_\phi)}{\pi})^2 \\
\Delta_t^{H}=-\tan \alpha \cot \beta[1.57-\frac{2}{3}\ln(\frac{M_H^2}{M_t^2})+\frac{1}{9}\ln(\frac{M_A^2}{M_t^2}))^2](\frac{\alpha_s(M_\phi)}{\pi})^2 \\
g_b^h = \frac{-\sin \alpha}{\cos \beta(1+\Delta_b)}(1-\frac{\Delta_b}{\tan \alpha \tan \beta}) \\
g_b^H = \frac{\cos \alpha}{\cos \beta(1+\Delta_b)}(1+\frac{\Delta_b}{\cot \alpha \tan \beta}) \\
g_b^A = \frac{\tan \beta}{1+\Delta_b}(1-\frac{\Delta_b}{\tan^2 \beta}). 
\end{align}

The quantity $\Delta_b$ is an effective shift in the b-quark mass induced by SUSY QCD corrections\cite{Hall:1993gn,
Carena:1994bv,Pierce:1996zz,Carena2006}. It is computed from the
sum of two terms, $\Delta_b = \Delta_b^b + \Delta_b^t$. 
\begin{align}
\Delta_b^b = \frac{2\alpha_s}{3\pi} m_{\tilde g} \mu \tan\beta I(m_{\tilde b_1}, m_{\tilde b_2}, m_{\tilde g}) \\
\Delta_b^t =\frac{\alpha_t}{4\pi} A_t \mu \tan\beta I(m_{\tilde t_1}, m_{\tilde t_2}, \mu) \\
I(a,b,c) = -\frac{1}{(a^2-b^2)(b^2-c^2)(c^2-a^2)}(a^2 b^2\ln\frac{a^2}{b^2}+b^2 c^2\ln\frac{b^2}{c^2}+c^2 a^2\ln\frac{c^2}{a^2})
\end{align}
In the equations above $\alpha_t \equiv \lambda_t/(4 \pi)$, where $\lambda_t = 
\sqrt{2}m_t/v_2$ is the top Yukawa coupling, and $A_t$ is the trilinear stop coupling. $\Delta_b$ can be
a significant correction to the decay rate. One can see that $\Delta_b^b$ decreases the rate for a positive sign of $\mu$ and
increases the rate for negative $\mu$. The $\Delta_b^t$ term may be comparable in size or quite small, with a variable sign depending
on the trilinear coupling. In addition to the Higgs decay vertex, we include a $\Delta_b$ correction factor at the production vertex.

In order to evaluate the expressions above, as well as the Higgs mass itself, we use the program ISAJET, which implements one of 
several available SUSY-breaking schemes in terms of GUT scale input parameters, then computes the low scale SUSY masses based 
on running of the parameters according to the renormalization group equations \cite{Paige:2003mg}. We use the full width of the 
Higgs as computed by 
ISAJET, modified by the corrected $b\overline{b}$ decay width as indicated above. In general, the full width of the Higgs is small
enough in comparison to the Higgs mass to use the Narrow Width Approximation (NWA)\cite{hbbb}. For large masses and high 
$\tan \beta$ the NWA becomes slightly worse. In practice we use a full Breit-Wigner calculation for $bg \to b\phi^0 \to bbb$ with
the coupling of the two outgoing $b$s from Higgs decay set to an effective value which reproduces the $b\overline{b}$ decay width. 

We simulate the signal using MadGraph4~\cite{Madgraph,Helas,Madgraph2} 
to generate our matrix elements ($bg \to bb\bar{b}$) and then evaluate 
the total cross-section with a Monte Carlo program. We smear the momenta of the outgoing b-quarks with a Gaussian distribution parametrized by
\begin{eqnarray}
\frac{\Delta E}{E} = \frac{0.60}{\sqrt{E}} \oplus 0.03,
\end{eqnarray}
based on ATLAS estimates of detector effects \cite{ATLAS}.
We then impose the following cuts on our signal:
\begin{itemize}
\item $p_T > 70$ for all three jets, 
\item $|\eta| < 2.5$ for all three jets,
\item $\Delta R_{ij} > 0.7$ for each pair of jets (i,j),
\item Missing $E_T < 40$ GeV, 
\item $|M_{ij} - M_\phi| < 0.15 M_\phi$ for at least one pair of jets.  
\end{itemize}

In the list above $\eta$ is the pseudorapidity and $\Delta R \equiv \sqrt{\Delta \eta^2 + \Delta \phi^2}$ is
a measure of jet separation. We choose the $p_T$, $\eta$, and $\Delta R$ cuts to provide for good reconstruction of
three jets in the b-tagging region. The $p_T$ cut is also chosen in accord with the CMS Level 1 Trigger for 3 jets \cite{CMS}. 
$M_{ij}$ is the invariant mass constructed from two of the three jets; we require that it be within a $15\%$ window of
the true pseudoscalar mass $M_A$ for at least one pair of jets. We found in a previous work that allowing only the two
highest $p_T$ jets for this requirement produces a slight improvement with respect to background at high masses \cite{hbbb}.

The backgrounds to our signal are dominated by pure QCD processes which produce three jets. The irreducible background is
$bg \to bb\bar{b}$. We also include the reducible backgrounds: $pp\to cbb + X$, $pp \to gbb +X$, and $pp \to qbb +X$ where
$q = u,d,s$ along with $pp \to t\overline{t} \to bbjjl\nu +X$ and 
$pp \to t\overline{t} \to bbjjjj +X$. These involve one mis-tagged
non-b jet and can potentially be reduced with improved b-tagging. For both the signal and the background we assume an effective 
b-tag rate of $\eps_b = 0.6$ with a mis-tag rate $\eps_c=0.14$ for c-quark jets and $\eps_j = 0.01$ for light jets ($g,u,d,s$).
With these efficiencies, the pure QCD backgrounds ($bbb,cbb,gbb,qbb$) are all of comparable size, with $bbb$ or $gbb$ being the 
largest, while the backgrounds with 
intermediate t-quarks are negligible in comparison. We include them nonetheless for completeness. The double mis-tagged background 
$ccb$ should be roughly $\frac{1}{16}$ the size of the $bbb$ background after tagging and thus only a $1-2\%$ correction to the total
background.

For simulation, backgrounds are generated using MadGraph amplitudes with the renormalization and factorization scales set to 
$p_T(1)/2$, half the transverse momentum of the leading jet, for the pure QCD backgrounds, and $\sqrt{s}$ for the $t\overline{t}$
backgrounds. Cuts are the same as used for the signal, except that we also apply a veto to $t\overline{t}$ events with 4 jets 
having $p_T > 15$ GeV. We assume a K factor of two for each background while keeping the K factor for the signal at one \cite{Vega,Bonciani}.

Once we have both signal and background we are able to compute a statistical significance, for which we use
\begin{align}
N_{SS} \equiv \frac{N_S}{\sqrt{N_S +N_B}}
\end{align}
where $N_S$ and $N_B$ are the number of expected signal and background
events respectively. We set $N_{SS} = 5$ as the discovery limit.
In practice, we use the following process to find the discovery contour: First we generate a set of background estimates over a range
of masses $M_A$. The background cross-sections only depend on the mass through the location and size of the invariant dijet mass
cut as given above. Next we choose a point in the GUT-scale parameter space for one of the SUSY-breaking models and use Isajet to
calculate the relevant weak scale parameters, namely, the Higgs mass and decay width. We modify the $\phi^0 \to b\overline{b}$ width
with the corrections listed above and feed these parameters into our MadGraph/Monte Carlo program which calculates our signal for 
that point in parameter space. The background at that point is determined by cubic spline interpolation from our array of $M_A$ 
dependent backgrounds and a significance is calculated. Then by scanning over one of the GUT-scale parameters with the others fixed
we can locate the discovery contour where $N_{SS} = 5$.

\section{ Experimental Constraints}

In addition to the estimated range of sensitivity in the 3$b$ channel, it is interesting to consider other effects of the high
$\tan \beta$ scenarios and model assumptions. There are a number of experimentally measured or constrained quantities which are
quite sensitive to deviations from the Standard Model. Especially important at high $\tan \beta$ is the rare decay $B_s \to 
\mu^+ \mu^-$. In the Standard Model this decay is expected to have the low branching fraction $BF(B_s \to\mu^+ \mu^-) =
(3.6 \pm 0.37)\times 10^{-9}$ \cite{Buras:2010mh}. However, diagrams involving $H$ and $A$ in a 2HDM are proportional to 
$(\tan \beta)^3$, meaning that their contributions to the decay rate scale as $(\tan \beta)^6$. This can change the predicted decay 
rate by orders of magnitude and the current results from several experiments put us in a very exciting time frame. 

At the Tevatron, D0 and CDF set limits on the observed branching fraction of $BF(B_s \to\mu^+ \mu^-) < 5.1 \times 10^{-8}$ and
$BF(B_s \to\mu^+ \mu^-) < 4. \times 10^{-8}$. Meanwhile, early results from the LHC have lowered this limit to $1.9 \times 
10^{-8}$(CMS) and $1.5 \times 10^{-8}$(LHCb)\cite{Chatrchyan:2011kr,lhcb-conf-2011-037}, using approximately $300 
\text{pb}^{-1}$ 
of data. A combined analysis using CMS and
LHCb data puts the limit at $1.1 \times 10^{-8}$\cite{lhcb-conf-2011-047}. Thus we are already constrained to limit the effects
of heavy Higgs bosons with large $\tan \beta$, while still allowing for an enhanced branching fraction up to 3 times the Standard 
Model rate. Interestingly, CDF has reported a weak excess corresponding to a signal at $(1.8^{+1.1}_{-0.9}) \times 10^{-8}$ with
$98\%$ probability of exceeding the SM rate \cite{Aaltonen:2011fi}. LHCb and CMS have not seen evidence of this signal but have also
not strongly ruled it out. LHCb in particular should cover the remaining space down to the Standard Model limit in the near future.
It was anticipated to reach the SM limit with $\sim 2 \;\text{fb}^{-1}$ of data at $14$ TeV running \cite{:2009ny}. Thus we expect it
to show evidence of an excess signal or to put stringent limits on the allowed branching fraction in the next year or two.

Supersymmetric models in general are constrained by electroweak precision data and searches for new flavor and CP violation. Two
constraints of particular interest to us are the measured values of $g_\mu-2$ and $b \to s \gamma$. Both quantities are sensitive
to $\tan \beta$ (though not so strongly as $B_s \to\mu^+ \mu^-$) and to the effects of the SM superpartners.

The anomalous magnetic moment of the muon, $g_\mu-2$, is one of the more precisely calculated and measured quantities of quantum
field theory. The experimental value is found to be $a_\mu \equiv (g-2)/2 = (116592089.0 \pm 6.3) \times 10^{-10}$ \cite{PDG 2010}.
A recent calculation puts the theoretical value for the Standard Model at $a_\mu  = (11659182.8 \pm 4.9) \times 10^{-10}$, leading
to a discrepancy $\Delta a_\mu = (26.1 \pm 8.1) \times 10^{-10}$, i.e. a $3.3 \sigma$ excess in experiment \cite{Hagiwara:2011af}.
Other groups find similar results. It is tempting to attribute this excess to new physics and supersymmetric models can easily
account for it if they have the correct set of masses and parameters. 

At high $\tan \beta$ the dominant new contribution to $a_\mu$ is proportional to $\tan \beta$ with the same sign as $\mu M_2$ \cite{Cho:2011rk}. (Unless $|M_2|, M_{\tilde E} \ll |M_1|,M_{\tilde L}$, where  $M_{\tilde L}, M_{\tilde E}$ are the left- 
 and right-handed slepton soft SUSY-breaking masses, respectively.)
Thus, for the minimal models we consider below, one requires a positive
sign for $\mu$ if MSSM contributions are to account for the observed excess of $a_\mu$. 

A third sensitive probe is the flavor changing decay $b \to s \gamma$, observed through $B \to X_s \gamma$. Experimentally, this is
measured at $(3.55 \pm 0.26) \times 10 ^{-4}$ \cite{PDG 2010}. Theoretical predictions in the Standard Model put the value at
$(3.15 \pm 0.23) \times 10 ^{-4}$ \cite{Buras:2011zb}.
 In supersymmetric theories, loops involving the charged Higgs boson, as well as those involving charginos
and squarks, can make large contributions to $b \to s \gamma$.

There are, of course, other precision constraints which one may take in to account. As we shall see in more detail, however, between
$B_s \to \mu^+ \mu^-$, $\Delta a_\mu$, and $b \to s \gamma$, the models we consider are already strongly constrained if we
wish to use them to fit the experimental data within reasonable error estimates.
We use the IsaTools \cite{Baer:1996kv,Baer:1997jq,Baer:2001kn,Mizukoshi:2002gs} set of subroutines incorporated with ISAJET to calculate our estimates of 
these observables.

\section{ Minimal Supergravity}

The first SUSY-breaking model we consider is minimal Supergravity mediation 
(mSUGRA)~\cite{
Chamseddine:1982jx,Ibanez:1982fr,Barbieri:1982eh,Hall:1983iz,Ohta:1983}. 
In the absence of other effects, gravity
should act as a messenger between the hidden sector where Supersymmetry is spontaneously broken and the Standard Model sector. 
That is, the scale of the messenger interactions $M_{mes}$ is approximately the Planck mass $M_{Pl}$, otherwise referred to as high-scale
SUSY breaking. This leads to a gravitino/goldstino with mass on the order of a few TeV. Gravitational interactions induce SUSY-
breaking terms at the high scale. 

In the minimal model, the GUT scale parameters are chosen to be 
a common scalar mass $M_0$, a common fermion mass $M_{1/2}$, 
a common trilinear coupling $A_0$, and the value of $\tan \beta$. 
All other parameters are fixed except the sign of $\mu$. 
We will consider only the $\mu > 0$ case, since otherwise we will 
have a larger than 3$\sigma$ discrepancy in $\Delta a_\mu$ 
for the decoupling limit and worse for detectable superpartners. 
The GUT scale parameters are run down to the weak scale, resulting in
a typical ratio of gaugino masses $M_1:M_2:M_3 \simeq 1:2:6$. 
The lightest supersymmetric partner is typically a neutralino.

In Fig.~1 we show $5 \sigma$ discovery contours (solid red) in the $M_0$,$M_{1/2}$ plane for four choices of $\tan \beta$ in
the mSUGRA model. We set $A_0 =0$ and $\mu > 0$. We present two contours, corresponding to 30 fb$^{-1}$ of data 
running at $14$ TeV and to 300 fb$^{-1}$ at that energy. For the lower luminosity we apply the cuts as described above. 
For the high luminosity figure, we assume more restrictive triggers will be required to reduce the total recorded event rate 
to manageable levels. For $\tan \beta = 30$ and higher, we raise the $p_T$ cuts to $150$ GeV\cite{ATLAS-thesis} and apply a reduced b-tagging 
efficiency $\eps_b = 0.5$. We also modify the invariant mass selection so that only the leading two jets in $p_T$ are considered 
as candidates for the Higgs mass peak. In our previous work we found that this strategy can improve the statistical significance 
and the signal to background ratio for high neutral higgs masses. For $\tan \beta = 20$, these cuts are very restrictive and 
would offer little improvement over the  $30 \; \text{fb}^{-1}$ results due to the relatively low masses accessible. We include a 
$300 \; \text{fb}^{-1}$ contour in the $\tan \beta = 20$ frame using a $p_T > 75$ GeV cut and $\eps_b = 0.5$ with any pair of jets 
considered as a candidate for the Higgs decay. 

 The dark gray
regions are excluded for theoretical reasons such as
tachyonic masses at the weak scale or lack of electroweak symmetry breaking. The solid blue region at low $M_{1/2}$ indicates 
charginos with mass $M_{\chi^+} < 103$ GeV, which have been excluded by experiment except in the case where $M_2 \gtrsim 1$ TeV or
sneutrino masses are less than $\sim 200$ GeV \cite{PDG 2010}. A more general bound including these cases can be put at $M_{\chi^+} > 92$ GeV. For
high values of $M_{1/2}$ and relatively low $M_0$, the lightest slepton, a stau, becomes the LSP. This region is indicated by the solid light-gray area on the plot.

The experimental value for $\Delta a_\mu$ is shown by the solid cyan line. The shaded cyan (forward slant hatched) region around it indicates a $\pm 2\sigma$ 
error around it.  We represent the experimental value and a $2\sigma$ error  for $b\to s\gamma$ with a green dash-dot-dot line
and yellow (backward slant hatched) shading.\footnote{We take the $1\sigma$ error to be the sum
in quadrature of the experimental and theoretical errors quoted above.} Note that for this choice of model the measured value (solid
 yellow) does not appear; the edge of the shaded region shows the
 lower $2\sigma$ limit. This is because SUSY contributions 
generate a negative correction to the SM value, while the experimentally measured value is slightly above the SM prediction. 
For $ B_s \to \mu^+ \mu^-$ we have drawn dashed magenta contours for $BF(B_s \to \mu^+ \mu^-) = 1. \times 10^{-8}$ and
$5. \times 10^{-9}$. The higher value, corresponding to the smaller area on the plot, is approximately the 
current exclusion limit, set by CMS and LHCb. It should be noted that this is also roughly the value
which provides the best fit for CDFs reported excess. The outermost line is not yet excluded by any experiment but should be reached
by LHCb in the near future.  LHCb should be able to approach the SM limit with a few $\text{fb}^{-1}$ of data, which in principle
would push the excluded region out to arbitrarily high SUSY masses, depending on the errors.

Current LHC data already strongly constrains some areas of parameter space. We include an 
exclusion bound based on LHC searches for SUSY particles in events with jets and missing energy \cite{arXiv:1109.2352}, this limit includes $1.1 \;\text{fb}^{-1}$ of data. It appears as 
as a blue, dash-dot line on the figure. However, the available bound was calculated based on a 
scenario with $\tan \beta = 10$ and so should not be taken as definitive here.
We also include a current exclusion bound for the mass of $A_0$ as a function of $\tan \beta$,based on LHC searches for the decay $H \to \tau^+ \tau^-$.
 \cite{htautau} This bound is 
shown by the dotted black line on the plot. At low $\tan \beta$ it does not significantly extend the excluded region beyond chargino searches, but at higher 
values a significant region corresponding to lighter pseudoscalar masses is already ruled out. It was calculated in an MSSM framework using the $M_H$ Max scenario and so may not exactly reflect the bounds  in the specific SUSY-breaking models we consider.
\begin{figure}[p]
\centering\leavevmode
\epsfxsize=5.8in
\epsfbox{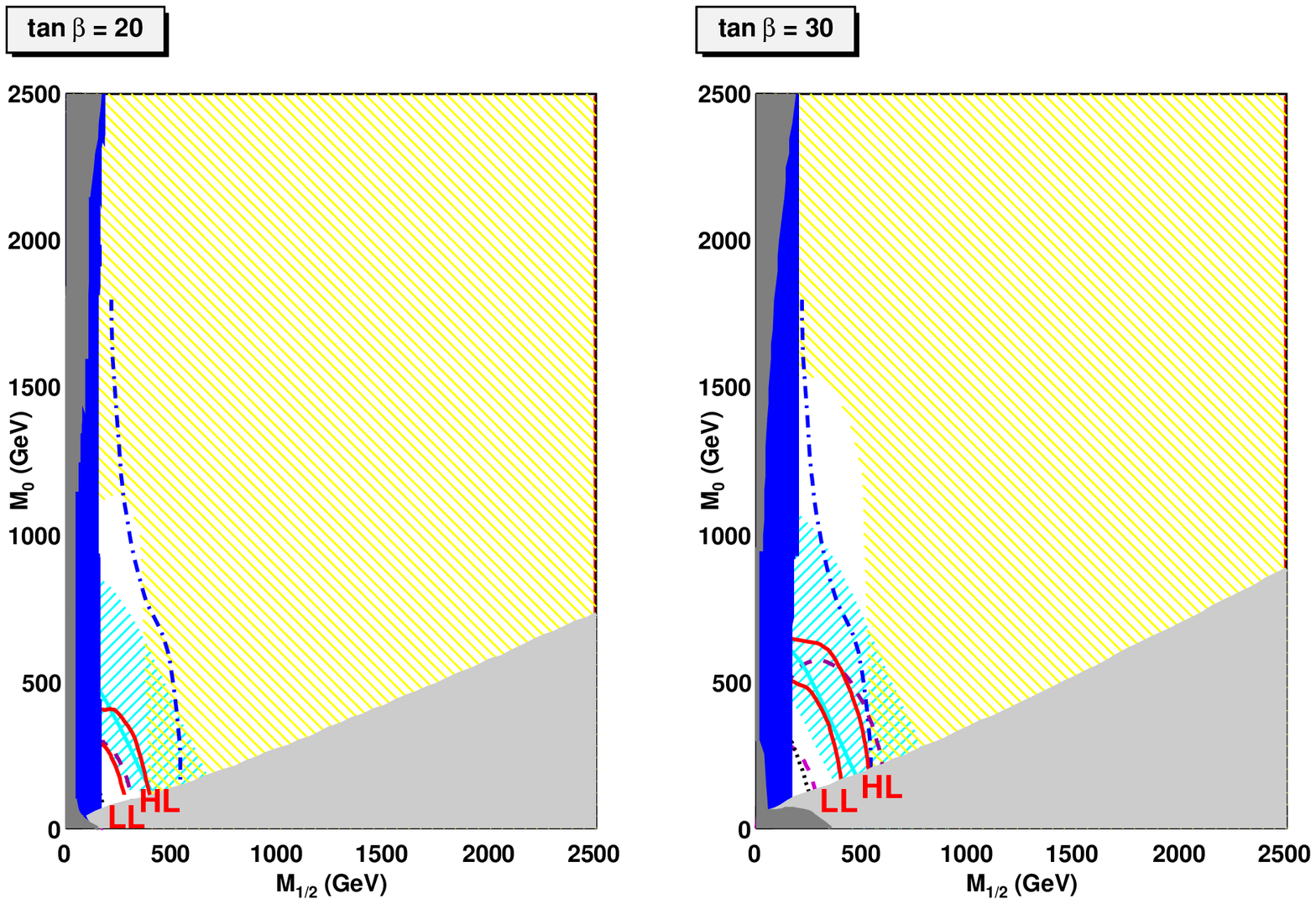}
\epsfxsize=5.8in
\epsfbox{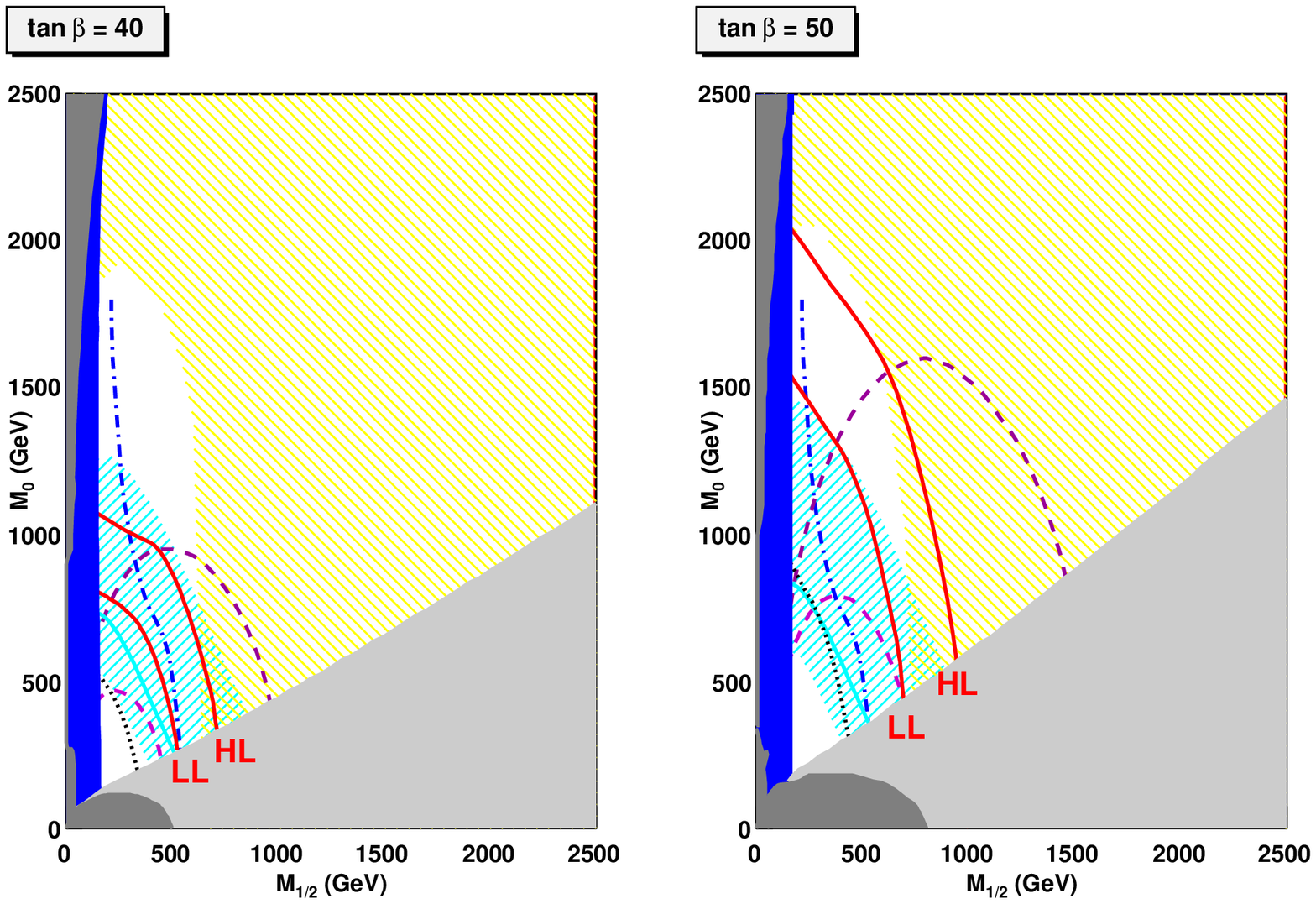}
\caption[]{
Discovery contours (solid red) for the $3$ $b$-quark Higgs signal in
mSUGRA with $30$ (LL) and $300$ fb$^{-1}$ (HL).
We have chosen $A_0 = 0$ and $\mu > 0$. 
Dark gray regions are excluded by theory. Light gray indicates a stau LSP. The dark blue area is ruled out by current 
chargino search limits. The experimental values with $2 \sigma$ errors are 
shown for $\Delta a_\mu$ (cyan, forward-slant hatched) and $b \to s \gamma$ (yellow, backward-slant hatched; central value in green dash-dot-dot). $B_s \to \mu^+ \mu^-$ limits of $1. \times 10^{-8}$ (current, lighter) 
and $5. \times 10^{-9}$ (darker) are indicated by dashed magenta lines. Current LHC exclusion limits are shown for $\phi^0 \to \tau^+ \tau^-$ (dotted black) and SUSY searches in jets plus missing $E_T$ (dash-dot blue).

\label{fig:sugra}
}
\end{figure}

Several comments are in order. First, we see that the discovery curve is roughly a quarter ellipse in $M_0$ and $M_{1/2}$. This 
approximately tracks the shape of a contour of constant $M_A$. An increase in  $M_0$ or $M_{1/2}$ tends to increase $M_A$ due to 
differential running of the Higgs parameters. Recall that in SUSY theories, the mass of $A^0$ is given to first order by
\begin{align}
   M_A^2 = \frac{M_{H_2}^2-M_{H_1}^2 -M_Z^2 \cos 2 \beta}{2 \cos 2 \beta}.
\end{align}
$M_{H_1}^2$ and $M_{H_2}^2$ are equal (set by $M_0$) at the GUT scale but differ at the weak scale due to different Yukawa and trilinear couplings that affect their running. $M_0$ and $M_{1/2}$ contribute to the size of these differences and a higher initial scale
$M_0$ can lead to larger differences after running. As we increase $\tan \beta$ we can detect heavier Higgs through the increased
Yukawa coupling to $b$s and the discovery contour moves to higher $M_0$ and $M_{1/2}$. For $\tan \beta \simeq 20$, this range is up to$M_0, M_{1/2} \sim 250-300$ GeV with $30 \;\text{fb}^{-1}$ of integrated luminosity. For $\tan \beta \simeq 50$, this extends $M_{1/2}$ out to $\sim 750$ GeV and $M_0 \sim 1600$ GeV. With 
high luminosity running this range is extended to $M_0, M_{1/2} \sim 350-400$ at low $\tan \beta$ and up to $M_{1/2}\sim 1000$ GeV or $M_0 \sim 2100$ GeV at $\tan \beta \simeq 50$.

As expected, the regions excluded by or soon to be explored by $ B_s \to \mu^+ \mu^-$ also grow rapidly with $\tan \beta$. For 
comparison one should bear in mind that the $ B_s \to \mu^+ \mu^-$ contours shown should be reached with only a few $\text{fb}^{-1}$
of data. The $ 30 \;\text{fb}^{-1}$ reach for the $3b$ signal significantly exceeds the current bounds from $B_s$ for all cases shown.
However, at high $ \tan \beta$ the $B_s$ search will rapidly begin to outperform the $3b$ range at high $M_{1/2}$ and relatively
low $M_0$. Conversely, $B_s \to \mu^+ \mu^-$ performance is limited for high $M_0$ and low $M_{1/2}$, even at high $\tan \beta$. 
The leading order terms in $\tan \beta$ are proportional to $M_{\chi^{\pm}}/M_H^{\pm}$ which becomes small in this region \cite{Bobeth:2001sq}.

The $2\sigma$ band around the measured value of $\Delta a_\mu$ prefers relatively light masses since we wish to generate
a significant non-zero value. This range gradually moves to higher mass values as we increase $\tan \beta$. For $\tan \beta \gtrsim 40$
the preferred range is almost entirely covered by a $3b$ Higgs search, while for lower values the lower $2\sigma$ region eludes us.
The measured value is covered by the $3b$ search for all cases shown.

A significant tension in these models is generated by the $b \to s \gamma$ prediction. The measured value is slightly in excess
of the SM prediction, while mSUGRA generically predicts a negative contribution to the decay rate. Thus only the lower $2\sigma$ 
bound appears on the plots and satisfying this constraint favors the decoupling limit with very high masses. The new physics 
contributions to $b \to s \gamma$ include Higgs-quark loops, which give a positive contribution, and chargino-stop loops
which unfortunately give us a larger negative contribution in this case.

The result is that if we wish to satisfy the measured $b \to s \gamma$ and $\Delta a_\mu$ values within $2\sigma$ errors on both
then we are pushed to a small region of parameter space in mSUGRA with $A_0$ and $sgn(\mu)$ as chosen. This overlap lies along the
stau LSP region with $M_{1/2} > 500$ GeV and low $M_0$. It is well known that the stau coannihilation region, where the lightest neutralino and the stau are nearly degenerate, provides one of the viable explanations for the observed dark matter relic density.  Statistical fit analyses \cite{Buchmueller:2011sw,Fowlie:2011mb,Bertone:2011nj} over the 
parameter space, taking into account a large number of constraints including dark matter 
density and searches, favor a similar region as indicated on our plots. In general, for this model, the $\Delta a_\mu$ constraint favors lighter SUSY mass parameters, while LHC data pulls us towards higher values. Large values of $\tan \beta$ are favored because they partially ameliorate this tension. 

The preferred area is covered by the $3b$ search with $ 300 \;\text{fb}^{-1}$ for 
$\tan \beta \gtrsim 40$ but becomes more difficult to cover with lower $\tan \beta$. It should, however, be well-explored by $B_s$ decay
in the near future for moderate to high $\tan \beta$. Inclusive direct searches for SUSY
particles also have the potential to rule out or favor this region with accumulating LHC data \cite{Baer:2009dn,Baer:2003wx}. With $100 \;\text{fb}^{-1}$, the LHC is expected to probe mSUGRA space up to 
$M_{1/2} \sim 1400$ GeV at low $M_0$ and $M_{1/2} \sim 700$ GeV at high $M_0$.

\section{ Anomaly Mediation}

If tree-level soft SUSY-breaking terms arising from supergravity are suppressed, there remain loop-level contributions arising
from the superconformal anomaly \cite{Randall:1998uk,Giudice:1998xp}. Such suppression can happen in extra-dimensional models 
where SUSY breaking does not occur on the brane which includes the SM sector. These anomalies generate mass terms which depend on 
the renormalization group beta functions and a mass scale set by the gravitino, $M_{3/2}$. 
\begin{align}
M_{\lambda i} = \frac{\beta_i}{g_i}M_{3/2} \\
M_{\phi}^2 = -\frac{1}{4}(\frac{\partial \gamma}{\partial g}\beta_g +\frac{\partial \gamma}{\partial y}\beta_y)M_{3/2}^2 \\
A_y = -\frac{\beta_y}{y}M_{3/2} \, .
\end{align}
The resulting SUSY spectrum at low scales
is significantly distinct from that found in mSUGRA. In particular, the gaugino masses have the ratio $M_1:M_2:M_3 \simeq 2.7:1:7.1$ 
where the gluino mass term has the opposite sign compared to the other two. This results in the lightest neutralino being primarily a
wino, with the lightest chargino and neutralino nearly degenerate. This has important consequences for decay phenomenology, with a long lived 
chargino. The purely anomaly generated terms are renormalization group invariant. However, this leads to an important problem: it
predicts tachyonic masses for the sleptons. To ameliorate this problem several solutions have been proposed which can generate
positive mass contributions from, e.g., bulk terms, gauge-mediated terms, new Yukawa terms, or higher order effects \cite{
Randall:1998uk,Pomarol:1999ie,Chacko:1999am,Katz:1999uw}.

 The minimal anomaly mediated model, mAMSB, is a simple
phenomenological model which assumes a universal additional mass term $M_0$, which is added to the anomaly generated scalar terms
at the GUT scale \cite{Feng:1999hg,Gherghetta:1999sw}. 
The addition of this new non-anomaly mediated term breaks the RG 
invariance of the soft-SUSY masses. The deviation from the pure-anomaly case for scalars depends at first order on the Yukawa
couplings. Hence, for first and second generation scalars the mass terms can remain close to their GUT-scale, diagonal relations 
and this represents a possible solution to the SUSY flavor problem. The complete set of GUT-scale parameters for the mAMSB model
can be given by $M_0, M_{3/2},\tan \beta, sgn(\mu)$. 
\begin{figure}[p]
\centering\leavevmode
\epsfxsize=5.8in
\epsfbox{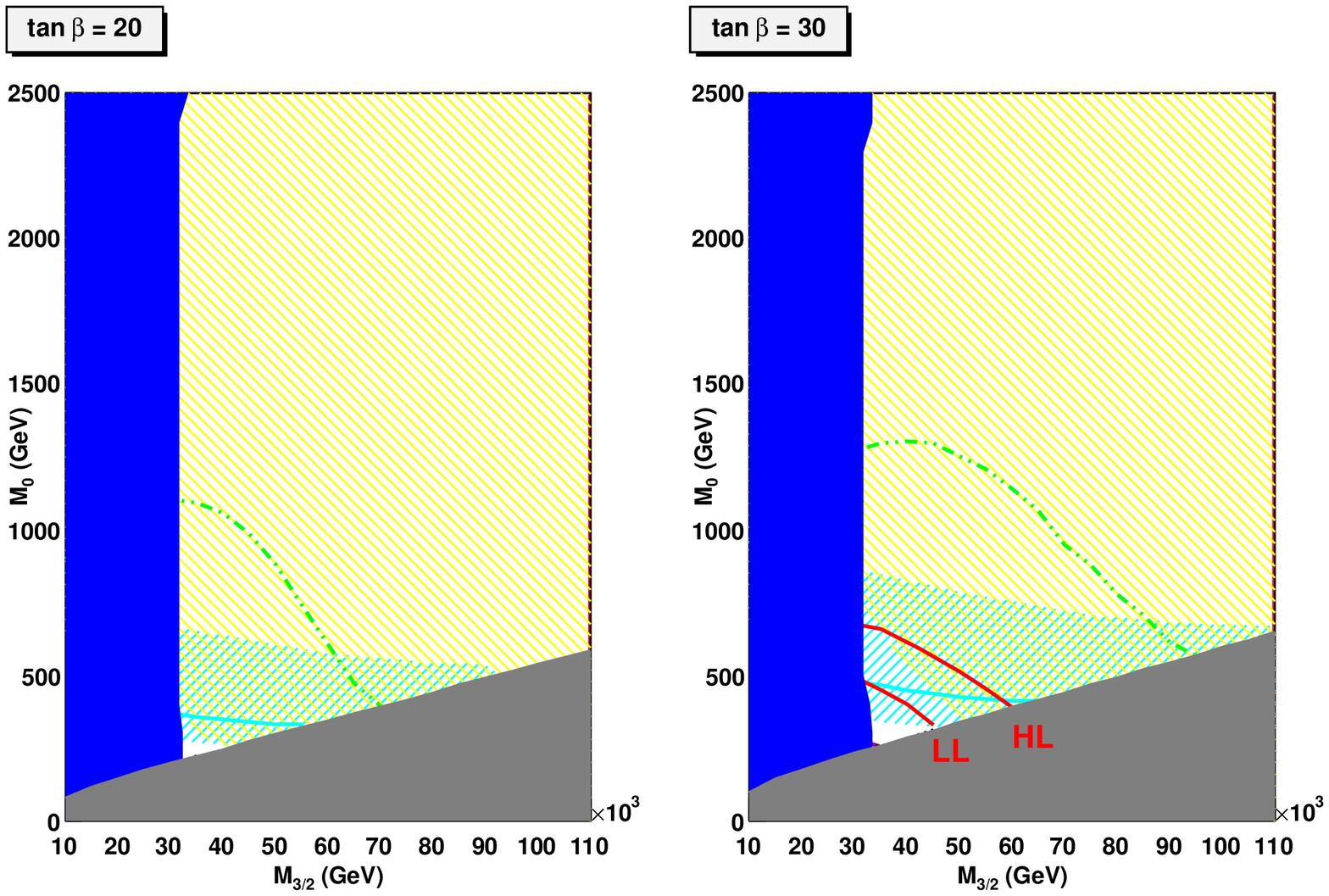}
\epsfxsize=5.8in
\epsfbox{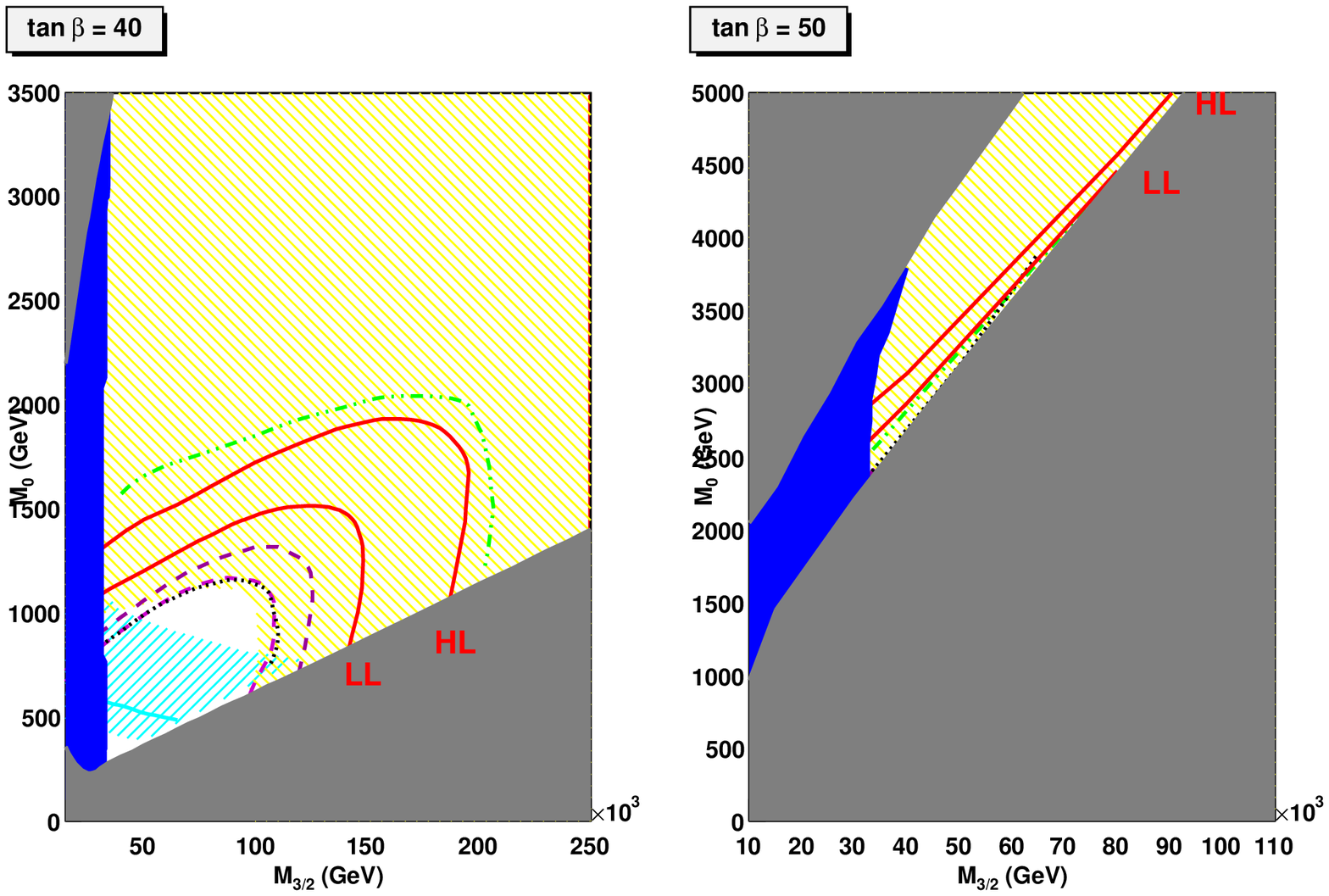}
\caption[]{
Discovery contours (solid red) for the $3$ $b$-quark Higgs signal in 
mAMSB with $30$ (LL) and $300$ fb$^{-1}$ (HL). We set $\mu > 0$. Dark gray regions are excluded by theory. The dark blue area is ruled out by current 
chargino search limits. The experimental values with $2 \sigma$ errors are 
shown for $\Delta a_\mu$ (cyan, forward-slant hatched) and $b \to s \gamma$ (yellow, backward-slant hatched; central value in green dash-dot-dot). $B_s \to \mu^+ \mu^-$ limits of $1. \times 10^{-8}$ (current, lighter) 
and $5. \times 10^{-9}$ (darker) are indicated by dashed magenta lines. Current LHC exclusion limits are shown for $\phi^0 \to \tau^+ \tau^-$ (dotted black).

\label{fig:amsb}
}
\end{figure}

In Figure 2 we show the $3b$ discovery curves in the $M_0, M_{3/2}$ plane for the four values of $\tan \beta$, as in the mSUGRA case. 
The color and pattern coding are the same as used before. The tachyonic region occurs where $M_0$ is not large enough to offset the 
negative contributions to scalar mass coming from $M_{3/2}$. This region grows as $\tan \beta$ is increased. As before, the discovery
curves roughly parallel curves of constant $M_A$, which exhibit interesting behavior for large $\tan \beta$. The mass depends on the
splitting between $M_{H1}$ and $M_{H2}$ at the weak scale and for low to moderate $\tan \beta$ the curve behaves similarly to
the mSUGRA case. Increasing $M_0$ or $M_{3/2}$ tends to increase $M_A$ due to the differential running and the starting point 
at the GUT scale. However, as can be seen in the plot for $\tan \beta = 40$, the situation becomes more complicated as the down-type
Yukawa couplings increase.

This is because, for values of $\tan \beta$ in this range, we approach an approximate Yukawa unification where the couplings of the
up and down-type quarks are very similar at the GUT scale. This does not occur with mSUGRA assumptions even for high $\tan \beta$ where
the weak scale coupling become similar. Threshold corrections at the SUSY scale are generically  large for the b-quark, whereas
they are required to be relatively small for good Yukawa unification \cite{Tobe:2003bc}. Following Wells and Tobe, the leading 
finite threshold corrections for the b-quark go as
\begin{align}
\delta_b \equiv \frac{y_b^{MSSM} - y_b^{SM}/\sin \beta}{y_b^{SM}/\sin \beta} \simeq \frac{-g_3^2 \mu M_{\tilde g} \tan \beta}
{12 \pi^2 M_{\tilde b}^2} + \frac{y_t^2 A_t \mu \tan \beta}{32 \pi^2 M_{\tilde t}}.
\end{align}
These corrections should be roughly $5\%$ or less for good unification, but for TeV squarks each term can be $\sim  50\%$ in general. For the mSUGRA plots 
shown above $A_t = 0$ at the GUT scale and becomes negative at the weak scale, thus both terms above are of the same sign and the 
correction is generically large enough to spoil any unification. In mAMSB on the other hand, $A_t$ is generated by the beta function
as shown above and is positive at the GUT and weak scales. It's size is also large enough at the weak scale to be comparable to the
gluino mass. The result is that it can significantly offset the corrections proportional to $g_3^2$, rather than add to them. This 
cancellation is not necessarily so precise as to guarantee exact Yukawa unification, but the difference between $y_t$ and $y_b$ at
the GUT scale is $\sim 10\%$ rather than $\sim 100\%$ as in mSUGRA. 

The result of this is that the running of $M_{H1}$ and $M_{H2}$ can become very similar in mAMSB for high $\tan \beta$ and secondary
effects become important. At low $M_{3/2}$, increasing $M_{3/2}$ tends to make the Yukawa
unification better,  generating less difference between the Higgs mass terms during
the running from the GUT to the weak scale and thus producing a lighter $A_0$. However, at 
the GUT scale the mass terms have a common positive contribution from $M_0$ and a negative
contribution which scales as $M_{3/2}$. Due to inexact Yukawa unification and contributions
to $M_{H1}$ from the tau coupling, the beta functions for $M_{H1}$ and $M_{H2}$ retain 
some difference and larger $M_{3/2}$ increases this initial difference. Hence, even with
very similar running the initial splitting becomes large enough that $M_{3/2}$ begins
to increase $M_A$. When $M_{3/2}$ is large and Yukawa couplings are similar, $M_A$ is not
particularly sensitive to $M_0$ since it does not contribute to any initial splitting
at the high scale. One can see that at the right edge of the curve $M_A$ actually increases
as $M_0$ is lowered toward the tachyonic bound. This is because the lepton contributions 
to the initial splitting and the differential running can become important. For high 
$M_{3/2}$ they can be negative and their magnitude increases as the offset $M_0$ decreases.

For $\tan \beta =50$, a detectable $M_A$ can correspond to very high values of both
$M_0$ and $M_{3/2}$ because of the effects discussed above. On the other hand, as can
be seen in the figure, the tachyonic region becomes very large, while another disallowed
region, one without EW symmetry breaking, grows for high $M_0$ and lower $M_{3/2}$. This
leaves only a narrow band of theoretically allowed parameter space.

As in mSUGRA, $\Delta a_\mu$ prefers relatively light superpartners which sit in a band at
low $M_0$ and $M_{3/2}$. This band is only moderately dependent on $\tan \beta$. The regions probed by
$B_s \to \mu^+ \mu^-$ grow quickly with $\tan \beta$ and current exclusion limits begin to become important
for $\tan \beta > 30$. The shape of the $B_s$ contour echoes that of the mass and discovery curves. Similar 
behavior  is also seen for the $b \to s \gamma$ allowed region. The situation with respect to $b \to s \gamma$ is, however,
notably different than in mSUGRA models. For mAMSB, the chargino-squark loops give a net positive contribution to
the decay rate. The additional Higgs contributions remain positive as before so the total correction relative to the 
Standard Model is the right sign to account for the observed value.

Qualitatively, this change in sign can be understood as a result of the $A_t$ term. If we consider $b$ to $s$ transitions with
each quark coupling to one side of a chargino-squark loop, the dominant terms come from the Higgsino-like coupling for the 
bottom quark paired with the Wino or Higgsino coupling of the strange quark. Before mass diagonalization, a left handed quark
couples to a left squark via the Wino or to a right squark via the Higgsino. Thus we have one term proportional to the 
Higgsino-Higgsino term in the chargino mass matrix times the left-right coupling in the squark mixing matrix. The other term
is proportional to the Higgsino-Wino mixing times the left-left or right-right coupling of the squarks. Most of these
remain qualitatively similar between the mSUGRA and mAMSB scenarios, however the off diagonal stop mixing term can 
change signs between the two. Recall that this term is generically given by
\begin{align}
M_{LR} \propto A_t - \mu \cot \beta.
\end{align}
For mSUGRA as shown above, we assumed $A_0 =0$ at the GUT scale and renormalization effects drive it to negative values at the weak
scale, so the mixing term is always negative for positive $\mu$. In
mAMSB, on the other hand, the $A_t$ term as given by Eq. ($20$) 
is positive at the GUT scale and remains positive and large enough at the weak scale to make $M_{LR}$ positive. The  sign
of the loop involving the Higgsino-Higgsino coupling is therefore changed while the other loops remain the same. As mentioned 
before, a generic feature of AMSB is a nearly pure Wino for the lightest chargino and, by extension, a pure Higgsino heavy 
chargino. Thus the loops which have changed sign tend to dominate over those which do not: the latter depend on the chargino 
mixing angle which is typically small. The end result is that the stop-chargino terms can give a moderate positive contribution to 
$b \to s \gamma$, unlike mSUGRA. Examining the figure we see that $b \to s \gamma$ now disfavors superpartners and Higgs which 
are too light because they would give too much positive correction to the Standard Model. For $\tan \beta \simeq 20-30$ we do
somewhat better than mSUGRA. With relatively low $M_0$ and high $M_{3/2}$ we can satisfy both $\Delta a_\mu$ and $b \to s \gamma$ 
within one or two $\sigma$ error on either measurement. As we push to higher $\tan \beta$, the tension increases as $b \to s \gamma$ requires higher masses while $\Delta a_\mu$ still favors lower values of the GUT scale parameters. Only two small wedges 
remain at $\tan \beta = 40$ where both predictions can satisfy the experimental numbers with $2\sigma$ errors on each. These 
wedges can be excluded by $B_s \to \mu^+ \mu^-$ in the near future and are well within reach of the $3b$ search. 

Based on current LHC data, exclusion bounds for mAMSB have been extrapolated in Ref. ~\cite{Allanach:2011qr}. 
The authors find that mAMSB is excluded for $M_{3/2} \lesssim 40$ TeV when $M_0 \lesssim 1$ TeV, for $\tan \beta 
= 10$.

At  $\tan \beta \simeq 20$ even the $300 \;\text{fb}^{-1}$ search does not cover a significant region beyond that excluded by chargino mass bounds. 
However the detectable space grows quickly with  $\tan \beta$. If $\tan \beta \simeq 30$ the $3b$ search probes a portion of the favored area 
with $300 \;\text{fb}^{-1}$, while $B_s$ decay does not currently 
put any new constraints on the parameter space. As $\tan \beta$ increases $30 \;\text{fb}^{-1}$ quickly becomes sufficient to explore the experimentally 
preferred areas. 
For $\tan \beta \simeq 50$ our search can probe into the theoretically
allowed band, but this solution is strongly disfavored by $\Delta a_\mu$. 
Direct searches for gluino and squark cascade decays at LHC 
with 100 fb$^{-1}$ are expected to probe to $M_{3/2}\sim 140$ TeV for 
low $M_0$ and to $M_{3/2} \sim 100$ TeV for high $M_0$ \cite{lhc_amsb}.
 
\section{ Gauge Mediation}

A third proposal for SUSY-breaking involves additional matter which is charged under the SM gauge interactions\cite{gmsb1,gmsb2,gmsb3}. If this matter 
also interacts with the hidden sector then it can act as the messenger for breaking. The minimal model, mGMSB, assumes N pairs of
fundamental $SU(5)$ GUT representations, $5 + \overline5$. In general, the gravitino mass is given by
\begin{align}
M_{\tilde G} = \frac{F}{\sqrt{3} M_{Pl}},
\end{align}
where F is the characteristic scale of SUSY breaking in the hidden sector. Since the superpartner masses are given schematically
by $\tilde M \sim F/M_{mes}$, the gravitino will be the LSP if the messenger mass, $M_{mes}$ is much below the Planck scale. 
This is the default assumption for mGMSB. The mass of the messenger 5-plets is an input to mGMSB models, but the superpartner 
masses only depend logarithmically on it. The observable SUSY masses are parametrized by $\Lambda \equiv \frac{F_s}{M_{mes}}$, 
where $F_s = F/C_G$. $C_G$ is introduced since the scale of SUSY breaking in the visible sector may be lower than that in the 
hidden sector. The complete set of high scale parameters is then $\Lambda, M_{mes},N,C_G,\tan \beta$ and $sgn(\mu)$. The scalar 
and gaugino masses are given by
\begin{align}
M_{\phi}^2 = 2N\Lambda^2(\frac{5}{3}(\frac{Y}{2})^2(\frac{\alpha_1}{4\pi})^2 +C_2(\frac{\alpha_2}{4\pi})^2+C_3(\frac{\alpha_3}{4\pi})^2) \\
M_{\lambda}=k_a N \Lambda \frac{\alpha_a}{4\pi},
\end{align}
where $k_1=5/3, k_2=k_3=1$, and Y is the hypercharge. $C_2 = 3/4$ for weak doublets and zero for singlets, $C_3 =4/3$ for squarks and 
zero otherwise. Changing the number of messenger fields N introduces a relative splitting between the 
gauginos and sparticles since the former scale as N and the latter as $\sqrt{N}$. Since the messenger scale can be well below the Planck 
scale and the presumed scale of flavor physics, and since the initial masses depend only on gauge couplings, GMSB is a potential 
solution to the SUSY flavor problem.

In the figures below we set $C_G = N =1$ and plot in $\Lambda, M_{mes}$.  The 
dark gray excluded region on the left comes from the requirement that $M_{mes} > \Lambda$. As expected, varying $M_{mes}$ has only a weak 
effect on most of the lines. For fixed $\Lambda$, increasing $M_{mes}$ has little effect on the overall scale of SUSY partners. This is 
seen in the contours for $\Delta a_\mu$, which are nearly flat. The SUSY corrections to $a_\mu$ come from chargino-sneutrino loops and 
neutralino-smuon slepton loops. They scale as $\sim \tan \beta M_\mu^2 / M_{SUSY}^2 $and are thus insensitive to changing $M_{mes}$ \cite{Endo:2001ym}. Increasing 
$M_{mes}$ does gradually tend to increase the pseudoscalar Higgs mass due to increased running effects.
\begin{figure}[p]
\centering\leavevmode
\epsfxsize=5.8in
\epsfbox{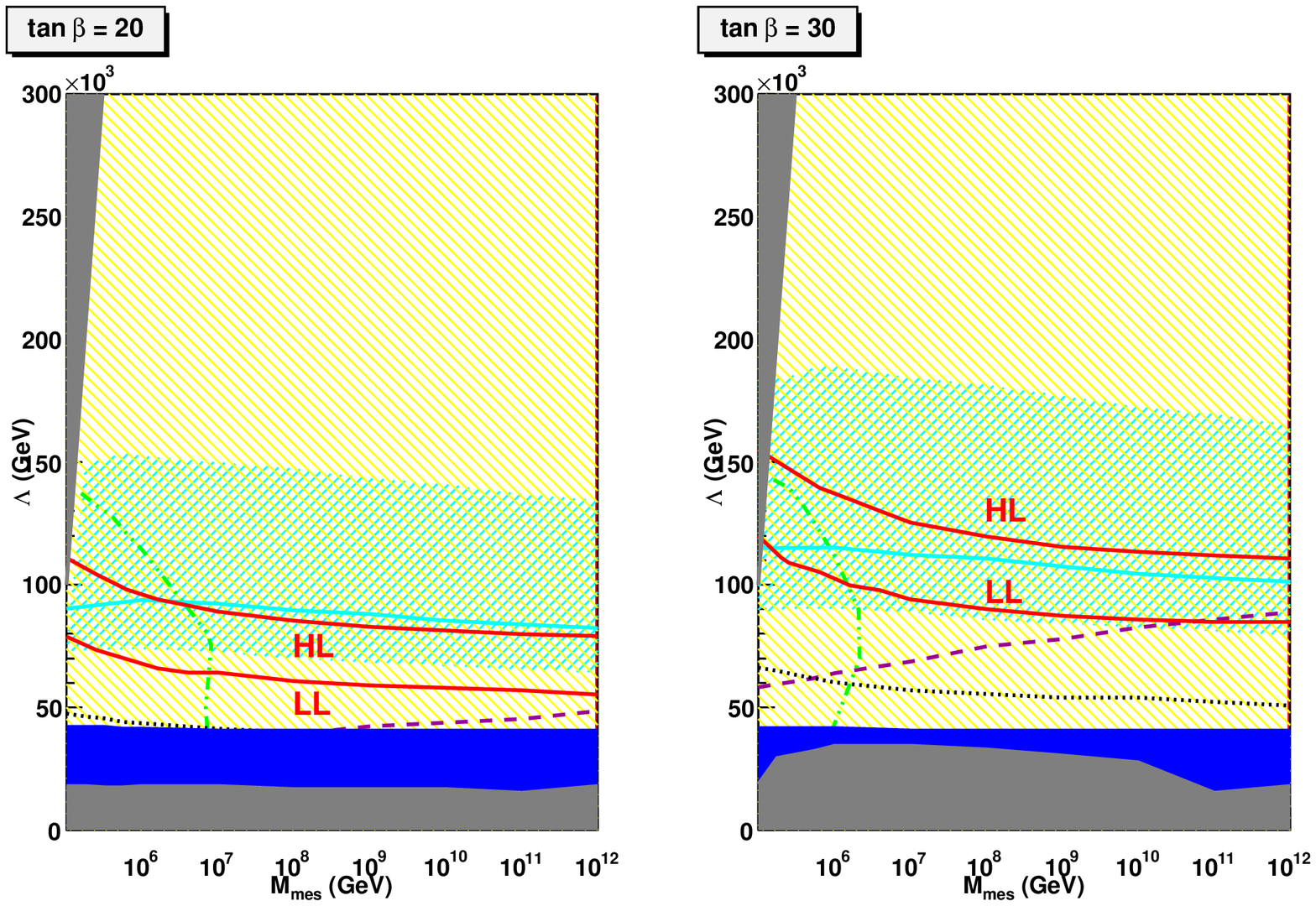}
\epsfxsize=5.8in
\epsfbox{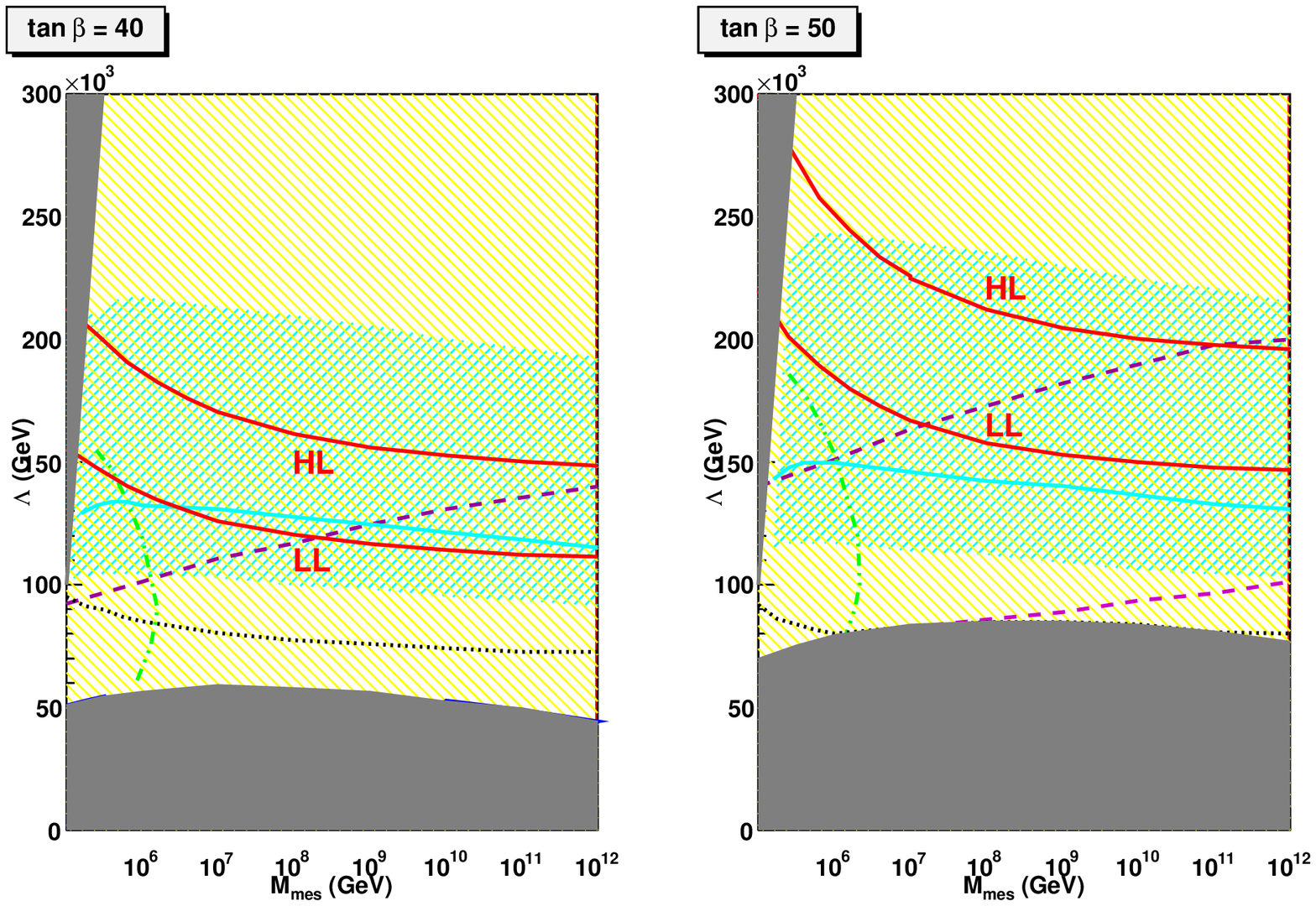}
\caption[]{
Discovery contours (solid red) for the $3$ $b$-quark Higgs signal in 
mGMSB with $30$ (LL) and $300$ fb$^{-1}$ (HL). We set $N = C_G =1$ and $\mu > 0$. Dark gray regions are excluded by theory.  The dark blue area is ruled out by current 
chargino search limits. The experimental values with $2 \sigma$ errors are 
shown for $\Delta a_\mu$ (cyan, forward-slant hatched) and $b \to s \gamma$ (yellow, backward-slant hatched; central value in green dash-dot-dot). $B_s \to \mu^+ \mu^-$ limits of $1. \times 10^{-8}$ (current, lighter) 
and $5. \times 10^{-9}$ (darker) are indicated by dashed magenta lines. Current LHC exclusion limits are shown for $\phi^0 \to \tau^+ \tau^-$ (dotted black).

\label{fig:gmsb}
}
\end{figure}

The curve for $b \to s \gamma$ exhibits some interesting behavior which requires a bit of explanation. For mGMSB, the predicted corrections 
to $b \to s \gamma$ are generically positive but not for the same reasons as in mAMSB. The $A_t$ terms are negative at the weak scale as in 
mSUGRA, so the contributions from chargino-squark loops are negative. However, they are comparatively small in mGMSB, so that the positive 
contributions from charged Higgs loops can dominate and the total correction is a small net positive. The chargino-stop terms are suppressed 
due to the super-GIM mechanism. That is, they depend on the squark mixing matrix which is unitary and in the limit of degenerate 
squark masses the various terms cancel as required for unitarity. For low scale SUSY breaking, running effects are small and the degeneracy 
of scalars with the same quantum numbers at the messenger scale is only mildly broken. As seen in the figure, increasing $M_{mes}$ at low 
scales decreases the contribution to $b \to s \gamma$ due to the increased Higgs masses and the spoiling of the squark degeneracy. Notably, 
although the overall correction to the SM value can become negative for large $M_{mes}$, it does not exceed the lower $2\sigma$ bound on 
the measured $b \to s \gamma$ rate.

For reasonably low $M_{mes}$ we can comfortably reconcile $\Delta a_\mu$ and $b \to s \gamma$ over a wide range of $\tan \beta$. $B_s \to 
\mu^+ \mu^-$ limits do not currently put a strong constraint on these favored regions although at high $\tan \beta$ they should begin to 
cover this region with enough luminosity. The preferred region also indicates excellent prospects for a $3b$ detection of the heavy neutral 
Higgs if $\tan \beta \gtrsim 30$. Much of the space would still be covered by this search even for somewhat lower $\tan \beta$. For a discussion of current mass limits see Ref.~\cite{Kats:2011qh}. The exclusion depends on the 
mass of the NLSP, which for our choice of parameters may be a neutralino or, at high $\tan \beta$ a stau. This 
puts a bound on the gluino mass of approximately $650$ GeV, which corresponds to $\Lambda \sim 80$ TeV.
Direct searches for sparticle pair production at LHC with 100 fb$^{-1}$ in the mGMSB model are expected to probe
to $\Lambda\sim 600$ TeV in the $\gamma\gamma +\eslt$ channel.\cite{lhc_gmsb}

\section{mSUGRA Revisited}

As discussed above, minimal supergravity appears to only marginally satisfy experimental constraints in a few limited regions of parameter space. 
This is due to the conflicting pull of the $g-2$ excess, which favors lighter SUSY partners, and the measured $b \to s \gamma$ branching fraction, 
which receives the wrong sign contribution from chargino loops in mSUGRA and therefore favors heavier masses. However, we have seen that mAMSB models 
can change the sign of the chargino contribution and that this can be traced back to the significant positive sign of $A_t$ at the weak scale. This 
suggests that by choosing a large, positive $A_0$ in mSUGRA we can induce a similar effect which may improve the prospects for this model.

In the figure below we plot results for mSUGRA as before but with $A_0 = 1.5$ TeV as our initial condition. This has several interesting effects on the 
weak scale observables. First we see that, as intended, the overall correction to $b \to s \gamma$ can be made positive and the experimental central 
value is predicted (solid yellow line) for $M_{1/2} \sim 200$ GeV with a weak dependence on $M_0$. The contours for discovery and for $\Delta a_\mu$ 
are not dramatically effected. On the other hand, the tachyonic region for lower values of $M_0$ and $M_{1/2}$ is notably larger than for the $A_0 = 0$ 
case. The stau LSP region also grows somewhat. The result is that it becomes possible to satisfy the $b\to s \gamma$ measured value while keeping the 
predicted $\Delta a_\mu$ within the $2\sigma$ lower bound.

 The region for $\Delta a_\mu$ exceeding the measured value is largely excluded by the requirement 
that we avoid tachyons. However, for $\tan \beta = 50$ a small area where $\Delta a_\mu$ falls exactly on the measured value remains. Moreover, the 
preferred value for $b \to s \gamma$ runs through this patch, so it is possible to have both  predictions very close to the measured numbers for 
$M_0 \simeq 700$ GeV and $M_{1/2} \simeq 200$ GeV.  One should note though, that the region within $2 \sigma$ error for $b \to s \gamma$ extends over 
the entire plot. It also appears that the area of exact overlap for $b \to s \gamma$ and  $\Delta a_\mu$ as currently measured may already be excluded by the $\phi \to \tau^+ \tau^-$ searches. As $M_{1/2}$ grows the chargino-squark corrections can again become negative but the SUSY effects diminish on the whole so we do not 
fall below the lower bound. 
\begin{figure}[p]
\centering\leavevmode
\epsfxsize=5.8in
\epsfbox{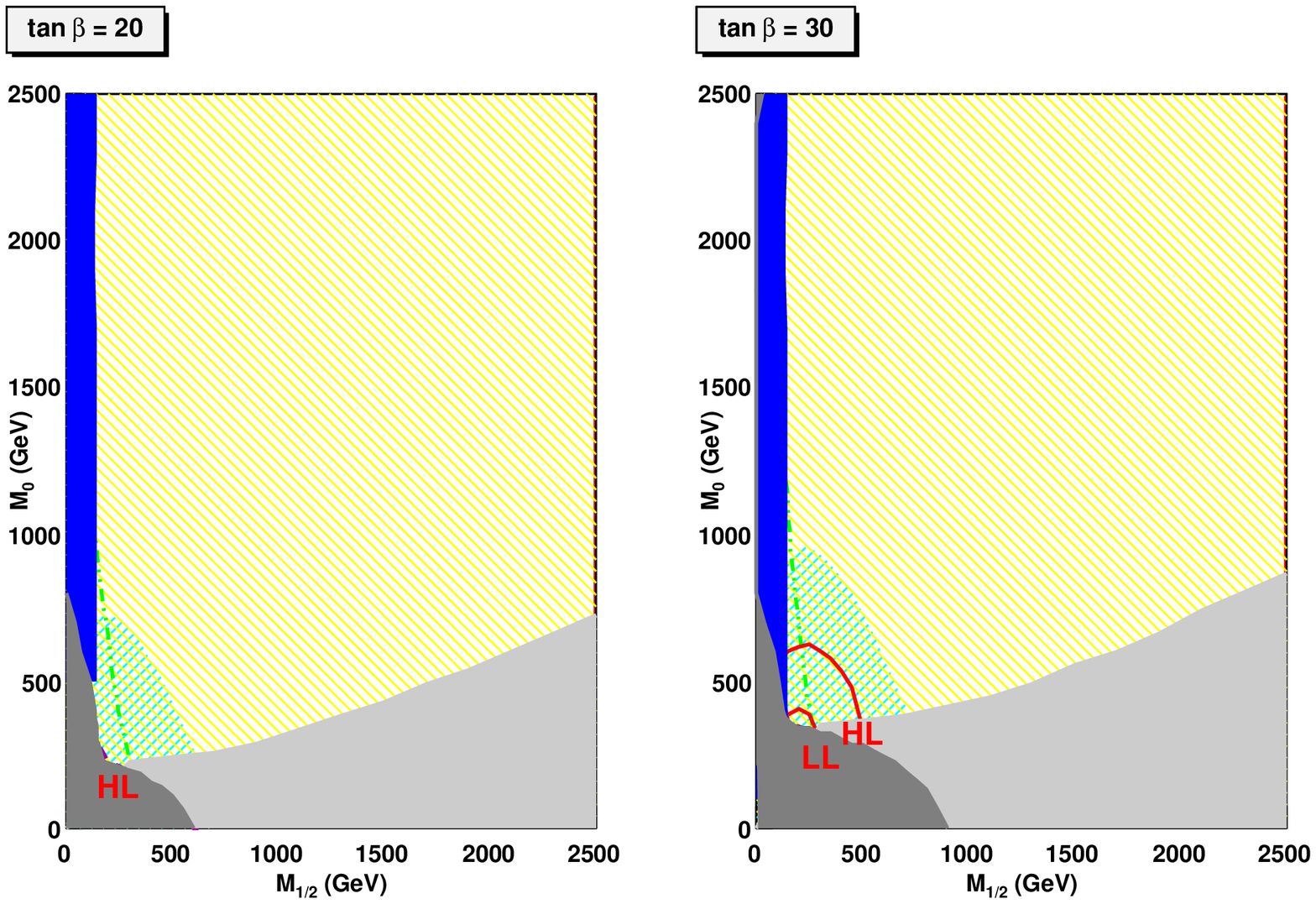}
\epsfxsize=5.8in
\epsfbox{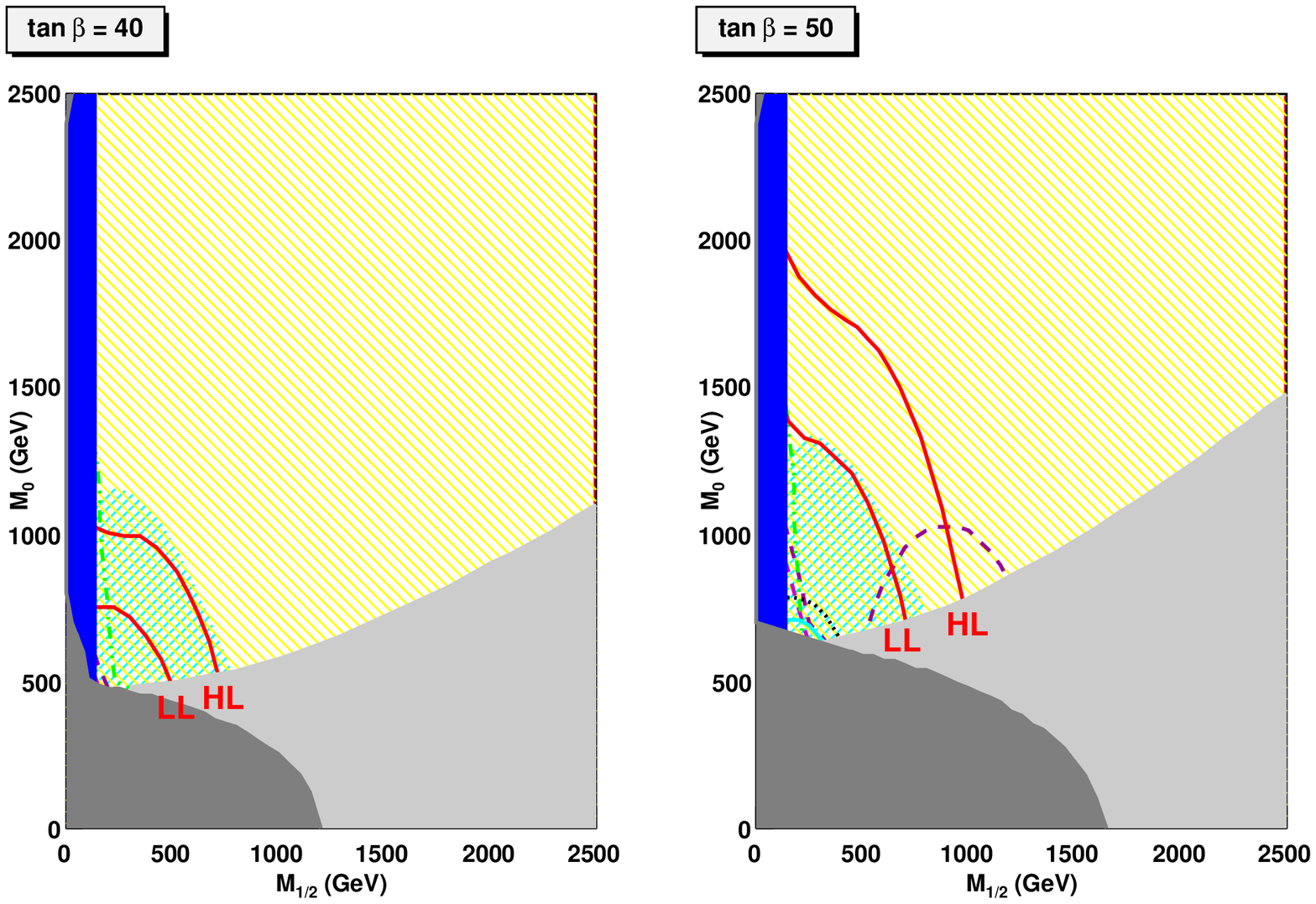}
\caption[]{
Discovery contours (solid red) for the $3$ $b$-quark Higgs signal in 
mSUGRA with $30$ (LL) and $300$ fb$^{-1}$ (HL). 
We have chosen $A_0 = 1.5$ TeV and $\mu > 0$. 
Dark gray regions are excluded by theory. Light gray indicates a stau LSP. The dark blue area is ruled out by current 
chargino search limits. The experimental values with $2 \sigma$ errors are 
shown for $\Delta a_\mu$ (cyan, forward-slant hatched) and $b \to s \gamma$ (yellow, backward-slant hatched; central value in green dash-dot-dot). $B_s \to \mu^+ \mu^-$ limits of $1. \times 10^{-8}$ (current, lighter) 
and $5. \times 10^{-9}$ (darker) are indicated by dashed magenta lines. Current LHC exclusion limits are shown for $\phi^0 \to \tau^+ \tau^-$ (dotted black).

\label{fig:msugra2}
}
\end{figure}

Another significant effect of positive $A_0$ is seen in the contours for $B_s \to \mu^+ \mu^-$. They are significantly reduced, such that only at $\tan 
\beta = 50$ does a limit of $BF < 5 \times 10^{-9}$ begin to cover parameter space that isn't already excluded by theoretical requirements. Additionally, the shape of the contours is slightly more complicated than the single ellipsoid seen with $A_0 = 0$. For $\tan \beta = 50$ the contours include a 
region at low $M_{1/2}$ which quickly fall offs for $M_{1/2} \sim 300$ GeV, then grows again to cover an ellipsoidal patch from $M_{1/2} \sim 500 - 
1400$ GeV. This can be qualitatively understood as follows: The ISAJET calculation of the branching fraction is based on an effective Lagrangian 
for flavor changing couplings between strange and bottom quarks \cite{Mizukoshi:2002gs}: 
\begin{align}
-\mathscr{L}_{eff} = \overline{D}_R f_DQ_l H_d + \overline{D}_R f_D[a_gM_g + a_uM_u f_u^\dagger f_u +a_wM_w]Q_L H_u^*
\end{align} 
where 
\begin{align}
a_g = -\frac{2\alpha_s}{3}\mu M_{\tilde g}, \quad a_u = -\frac{1}{16\pi^2}\mu A_t, \quad a_w = \frac{g^2}{16\pi^2} \mu M_2.
\end{align}
$M_g$,$M_u$ and $M_w$ are diagonal mass matrices that depend on loop functions arising from gluino-down squark, stop-higgsino, and wino-up squark
graphs respectively. The effective FCNC interactions are proportional to a function $\chi_{FC}$ which depends on a sum over these three 
terms with appropriate mixing coefficients. (See reference for details.) For our regions of interest the wino-squark loops are not as important 
as the other contributions. At low $M_{1/2}$ both the gluino-squark and higgsino-stop loops are of the same sign, corresponding to a positive 
$A_t$ at the weak scale and generating a significant enhancement of $B_s \to \mu^+ \mu^-$. As $M_{1/2}$ increases, the sign of $A_t$ at the 
weak scale changes due to running effects. For $M_{1/2} \sim 1000$ GeV the higgsino-stop term dominates as $A_t$ becomes large and negative, 
leading to a detectable excess in the branching fraction until the relevant particles become too heavy. However, for intermediate values of 
$M_{1/2}$ the $A_t$ terms are of the right order to cancel the other contributions, giving us the trough seen in the graph. 

In general, the discovery contours in this high $A_0$ scenario are similar to the $A_0$ case. One can see that they cover most of the 
experimentally preferred region for $\tan \beta \gtrsim 40$. This scenario is quite interesting to us, since it shows the possibility, 
particularly at high $\tan \beta$, to satisfy both the $b \to s \gamma$ and $\Delta a_\mu$ measurements in a region which is not 
particularly sensitive to $B_s \to \mu^+ \mu^-$ but should be well within the range of $3b$ (and related) searches.

\section{Conclusions}

The search for supersymmetry and the details of electroweak symmetry breaking is currently in an exciting phase with the LHC quickly 
accumulating data at 7 TeV and expected to increase energy in 2014. Updated constraints on precision measurements limit the parameter space 
in many models and some will rapidly improve with LHC data. 
A key signature of supersymmetry is the extended Higgs sector with two 
additional neutral particles and a pair of charged Higgs, unlike the Standard Model. 

To limit the rather large number of parameters in the MSSM, 
we have considered three simple models of SUSY breaking: mSUGRA, mAMSB, and mGMSB. 
We have delineated the parameters space regions of these models 
where a $3$ $b$-quark signature arising from $bA^0$, $bH^0$ production should be visible at LHC with 30 or 300 fb$^{-1}$ of
integrated luminosity. Since the Higgs to $b$-quark couplings become large at high $\tan \beta$ 
(a scenario favored  by Yukawa coupling unification), prospects for such a Higgs search are most promising at
large $\tan\beta$ values. The $3b$ signal will be complementary to direct sparticle searches, 
and will provide information on the heavy Higgs sector of the model.

The rare decay $B_s \to \mu^+ \mu^-$ is also highly sensitive to $\tan \beta$ and already strongly constrains 
new physics at very high $\tan \beta$ values. 
Limits from this decay mode are expected to improve quickly, potentially excluding important regions of parameter space. At 
the same time, CDF has seen a weak signal near the current limit which would strongly suggest new physics if it 
were to be confirmed. 

Meanwhile, the measured $3\sigma$ excess of $a_\mu$ suggests a positive sign for $\mu$ and relatively light SUSY particles. 
For mSUGRA, this is in tension with the experimental value of $b \to s \gamma$, which is somewhat above the SM prediction, 
while the theoretical prediction receives negative corrections from chargino-squark loops. 
This pulls one toward higher masses to minimize the corrections and only a small region 
of parameter space is left where both numbers can fall within $2 \sigma$ deviations. It is possible that a large, positive value for $A_0$ 
at the GUT scale can ameliorate this tension. For moderate to high $\tan \beta$, the $3b$ search will probe into the allowed region, as will
$B_s \to \mu^+ \mu^-$ in the near future.

Anomaly mediated models provide one way to improve the predictions, since they give a positive contribution to $b \to s \gamma$ for a 
positive sign of $\mu$. This solution favors moderate ($\lesssim 40$) values of $\tan \beta$ so as not to increase $b \to s \gamma$ by too 
much and to avoid the rapidly growing constraints from $B_s \to \mu^+ \mu^-$. The $3b$ search will cover most of the preferred space  
for scenarios with $\tan \beta > 30$ but will require a very high integrated luminosity to probe lower values. 

Perhaps the most natural fit to experiment, among the models we consider, is mGMSB. In this case, relatively light messenger fields lead 
to degenerate squarks which results in a cancellation of the negative chargino-stop terms in $b \to s \gamma$, leaving an overall small 
positive contribution. This allows one to fit both the $\Delta a_\mu$ and $b \to s \gamma$ measurements easily. If this is indeed a hint at 
the nature of SUSY-breaking then the $3b$ search is quite promising and $A^0/H^0$ should be discoverable at the LHC for a large range of 
moderate to high $\tan \beta$.  

\newpage


\end{document}